\documentclass[intlimits,twoside,a4paper]{article}
\usepackage{amsmath,amssymb}
\usepackage{graphicx}
\usepackage{wrapfig}
\usepackage[T2A]{fontenc}
\usepackage[cp1251]{inputenc}
\usepackage[eqsecnum]{cmpj2}

\issue{2012}{15}{4}{43701}

\doinumber{10.5488/CMP.15.43701}

\DeclareMathOperator{\Tr}{Tr}

\title[Exact solution of a variety of X-ray probes]%
{Exact solution of a variety of X-ray probes \\ in the Falicov-Kimball model with dynamical mean-field theory }
\author[A.M. Shvaika, J.K. Freericks]{A.M. Shvaika\refaddr{label1}, \
        J.K. Freericks\refaddr{label2}}
\addresses{
\addr{label1} Institute for Condensed Matter Physics of the National Academy of Sciences of Ukraine, \\
1 Svientsitskii St., 79011 Lviv, Ukraine
\addr{label2} Department of Physics, Georgetown University, 37th and O Sts. NW, Washington, DC 20057, USA
}

\date{Received July 11, 2012}
\authorcopyright{A.M. Shvaika, J.K. Freericks, 2012}

\begin{document}

\maketitle

\begin{abstract}
We examine the core-level X-ray photoemission spectroscopy (XPS),
X-ray absorption near-edge spectroscopy (XANES) and X-ray emission
spectroscopy (XES) in the Falicov-Kimball model by using the exact
solution from dynamical mean-field theory.  XPS measures the
core-hole propagator, XANES measures the absorption of X-rays when
the core electron is excited to an unoccupied electronic state of
the solid and is not emitted, and XES measures the spectra of
light emitted as electrons fill the core-hole state created via
some form of X-ray excitation. These three spectra are closely
related to one another and display orthogonality catastrophe
behavior at $T=0$. We show an efficient way of evaluating these
spectra at finite temperature, with a primary focus on the details
of XANES.
\pacs 71.10.--w, 71.27.+a, 71.30.+h, 02.30.Rz
\end{abstract}

\section{Introduction}

\looseness=-1
X-ray-based spectroscopy is one of the most important tools in
understanding condensed matter physics. Experiments in X-ray
spectroscopy are quite sophisticated and are capable of measuring
a large range of different properties. The basic idea, however, is
quite simple. X-rays are shone onto a sample and one measures
either photons or electrons that are emitted after the scattering.
Depending on what is measured, in what energy range, and how the
initial X-ray energy is tuned relative to different absorption
edges, one finds a ``zoo'' of different probes that are used
within X-ray scattering.  Here we focus on three of these probes,
assuming we are using monochromatic X-rays shining on our sample.
The first is X-ray photoemission spectroscopy (XPS) of the core
levels.  In this experiment, the X-ray is tuned far enough above
an absorption edge for the photoexcited core electron to have
enough energy to rapidly escape from the crystal (without
interacting with it) and to effectively leave the sample as a free
particle. One measures the rate at which the electrons hit a
detector as a function of the final energy of the electron,
integrating over all angles (if one measures this as a function of
the angle, then one can determine the angle-resolved photoemission
spectroscopy, but this is usually unimportant for core-level
spectra which have essentially no momentum dependence). In this
sense, it is a photon in, an electron out spectroscopy. The second
is X-ray absorption near edge spectroscopy (XANES) which measures
the attenuation of X-rays as they are shone onto a material as a
function of the incident X-ray energy. The X-ray energy is tuned
such that the photoexcited electron does not have enough energy to
leave the crystal, and hence it is similar to an optical
absorption measurement, but with a different frequency of light.
In order to get a spectrum, one needs to change the X-ray
frequency and measure the absorption as a function of the color of
the X-ray light, just like one measures the optical conductivity
(after processing the reflectivity data to extract the
absorption). If the electron does leave the crystal but with a
low-enough energy to strongly interact with the surface, then one
has an X-ray absorption fine structure (EXAFS) experiment, which we are not
going to discuss hereinafter. The third is X-ray emission
spectroscopy (XES), where one has a system with an excited
core-hole, and we measure the spectra of the light emitted from
the material as the conduction electrons fill the hole and cause
the sample to fluoresce. It is the stimulated emission of the core
hole back down to the ground state, which is enhanced due to the
presence of the X-ray radiation, and hence this is also an elastic
process where one measures the light emitted at the same frequency
as the incident light, but in different directions.

\looseness=-1
One can contrast these probes with inelastic scattering such as
nonresonant inelastic light scattering (NIXS) and resonant
inelastic light scattering (RIXS) which are coherent second-order
processes similar to Raman scattering at optical wavelengths. XPS
is a single-particle process that can be analyzed via the sudden
approximation, XANES and XES are first-order processes in the
matter-light interaction and correspond to two-particle processes,
while NIXS and RIXS are second-order processes in the matter-light
interaction which correspond to four-particle correlation
functions and will not be further discussed here (see~\cite{nandan_rixs} for further details in the Falicov-Kimball model).

\looseness=-1
Before we discuss the formal developments in detail, we need to define our model.  We work with the Falicov-Kimball
model~\cite{falicov_kimball}, with an additional core-hole interaction.  For simplicity, we work in a hole picture, where
the fermionic operators are all hole rather than particle operators. The spinless version of the Falicov-Kimball model  involves the interaction of itinerant holes (described by the creation operator $d_i^\dagger$ at site $i$) which can hop between nearest neighbors and localized
holes (described by the creation operator $f_i^\dagger$ at site $i$) which do not hop, but interact with the conduction
holes via a Coulomb interaction $U$.  When we add the core-hole (described by the creation operator $h_i^\dagger$ at site $i$), it interacts via an on-site Coulomb repulsion $Q_d$ with the conduction holes
and $Q_f$ with the localized holes. The Hamiltonian is then
\begin{align}
\mathcal{H}=&-\sum_{ij}t_{ij}d^\dagger_id^{}_j-\mu\sum_id^\dagger_id^{}_i+E_f\sum_if^\dagger_if^{}_i
+U\sum_id^\dagger_id^{}_if^\dagger_if^{}_i+E_h\sum_ih^{\dagger}_ih^{}_i\nonumber\\
&+\sum_i\left(Q_dd^\dagger_id^{}_i+Q_ff^\dagger_if^{}_i\right)h^\dagger_ih^{}_i
\label{eq: ham}
\\
= & \; \mathcal{H}^{\textrm{kin}}+\sum_i\mathcal{H}^{\textrm{loc}}_{i},
\end{align}
where $t_{ij}$ is a real symmetric matrix that is nonzero only if
$i$ and $j$ are nearest neighbors and it equals $t^*/2\sqrt{d}$ on
a $d$-dimensional hypercubic lattice (we will take the limit
$d\rightarrow\infty$ and use $t^*$ as the energy unit). The symbol
$\mu$ is the chemical potential for the conduction holes, while
$E_f$ and $E_h$ are the site energies for the localized hole and
core hole, respectively. The term $\mathcal{H}^{\textrm{kin}}$
denotes the first term of the Hamiltonian which includes the
kinetic energy of the conduction electrons, while
$\mathcal{H}^{\textrm{loc}}_i$ includes all of the remaining terms
at site $i$, which are the local terms. Since the localized hole
number operators $f^\dagger_if^{}_i$ and $h^\dagger_ih^{}_i$ both
commute with $\mathcal{H}$, they are conserved quantities, and
hence this model maps onto a generalized Falicov-Kimball model
with spinless conduction and localized holes,
which can be solved exactly within dynamical mean-field theory
(DMFT) as described in detail in the review in
reference~\cite{freericks_review}.

The retarded local Green's function for the conduction holes is defined via
\begin{equation}
G_d^{r}(t-t^\prime)=-\ri\theta(t-t^\prime)\left\langle \left\{d^{}_i(t),d^\dagger_i(t^\prime)\right\}_+\right\rangle,
\label{eq: green_c}
\end{equation}
where the hole operators are expressed in the Heisenberg picture $d^{}_i(t)=\exp(\ri\mathcal{H}t)d^{}_i\exp(-\ri\mathcal{H}t)$ and the angle brackets imply a trace over all states weighted by the density matrix $\rho=\exp(-\beta\mathcal{H})/\mathcal{Z}$ with $\mathcal{Z}=\Tr \exp(-\beta \mathcal{H})$ and $\beta=1/T$ the inverse temperature. The symbols $\{\cdot,\cdot\}_+$ denote the anticommutator. Since we have time-translation invariance in equilibrium, the Green's function depends only on the time-difference, and hence it is
convenient to take a Fourier transform and represent it as a function of frequency $\omega$. In addition, we
need to define the self-energy $\Sigma(\omega)$, effective medium $G_0(\omega)$, and dynamical mean field $\lambda(\omega)$. This is done as follows: (i) the local Green's function can be expressed in terms of a Hilbert transform of the self-energy with a weighting function given by the noninteracting density of states $\rho(\epsilon)=\exp\left(-\epsilon^2\right)/\sqrt{\pi}$ which is a Gaussian for the infinite-dimensional limit of a hypercubic lattice
\begin{equation}
G_d^{r}(\omega)=\int \rd\epsilon \rho(\epsilon) \frac{1}{\omega+\mu-\Sigma(\omega)-\epsilon}\,;
\label{eq: dyson_momentum}
\end{equation}
(ii) the effective medium is then defined via the Dyson equation for the local Green's function
\begin{equation}
\frac{1}{G_0(\omega)}=\frac{1}{G_d^{r}(\omega)}+\Sigma(\omega);
\label{eq: dyson_loc}
\end{equation}
and (iii) the dynamical mean field is then
\begin{equation}
\lambda(\omega)=\omega+\mu-\frac{1}{G_0(\omega)}\,.
\label{eq: lambda}
\end{equation}
It turns out that the dynamical mean field plays a prominent role in determining different X-ray spectroscopies.

Note that we are always working in the limit where the number of core holes in the system is so small, that the density
vanishes in the thermodynamic limit. Hence, in determining the dynamical mean field, we can set all $h^\dagger_ih^{}_i$
operators to zero and recover the spinless Falicov-Kimball model. The core-hole operators, however, are quite important when we calculate the X-ray spectroscopies. The details
of the numerical solution of the Green's functions, self-energies, and dynamical mean fields of the spinless Falicov-Kimball model can be found in reference~\cite{freericks_review}.

In the next section, we discuss the many-body correlation functions for different spectroscopies.  In section~3, we describe the exact solution of these correlation functions within the DMFT approach.  In section~4, we show some numerical results, and we conclude in section~5.

\section{Many-body formalism for different correlation functions}

\begin{wrapfigure}{r}{0.5\textwidth}
 \centerline{\includegraphics [width=2.0in, angle=0, clip]  {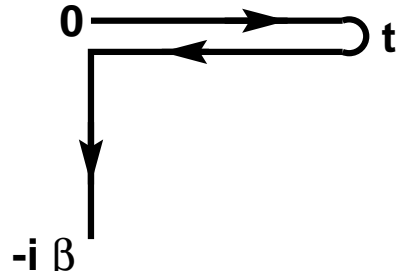}}
\caption[]{ The Kadanoff-Baym-Keldysh contour.  The contour starts at time 0, moves along the real time axis to time $t$, moves back along the real axis to time 0, then extends down the imaginary axis to imaginary time $-\ri\beta$.
}
\label{fig: contour}
\end{wrapfigure}

We start with a discussion of XPS.  This is a photoemission process.  If we invoke the sudden approximation and ignore the momentum dependence of matrix elements which connect the free electron states with the scattering states, then the XPS signal is simply given by a greater core-hole Green's function (see, for example, reference~\cite{freericks_sudden}). This Green's function can be found from the so-called contour-ordered Green's function which is defined for each time lying on the Kadanoff-Baym-Keldysh contour~\cite{kadanoff_baym,keldysh} depicted in figure~\ref{fig: contour}. The contour starts at time $\bar t=0$, runs out to time $\bar t=t$, then runs backwards in time and finally moves along the imaginary axis a distance of $\beta=1/T$. The greater Green's function has the first time lying on the lower branch and the second time lying on the upper branch, so the first time is ahead of the second one on the contour.


Before we define the  contour-ordered core-hole propagator, we need to extend the definition of the dynamical
mean field from real frequencies to times on the contour.  This is done by defining the following function
\begin{equation}
\lambda_c(\bar t,\bar t^\prime)=-\frac{\ri}{\pi}\int_{-\infty}^{\infty}\rd\omega \Im[\lambda(\omega)]
\re^{-\ri\omega(\bar t-\bar t^\prime)}\left [ f(\omega)-\theta_c(\bar t,\bar t^\prime)\right ],
\label{eq: lamda_c}
\end{equation}
where $f(\omega)=1/[1+\exp(\beta\omega)]$ is the Fermi-Dirac distribution function and $\theta_c(\bar t,\bar t^\prime)$
is the generalization of the Heaviside unit step function to the contour, which equals one if $\bar t$ is ahead of $\bar t^\prime$ on the contour, equals zero when $\bar t$ is behind $\bar t^\prime$ on the contour and equals 1/2 when $\bar t=\bar t^\prime$.  Note that on
the contour, two times which numerically have the same value are only considered equal if they also lie on the same branch
of the contour, and the designation ahead or behind refers to this ordering along the contour as illustrated by the arrows
in figure~\ref{fig: contour}.

The contour-ordered Green's function for the core hole is then defined to be $G_h^c(t,t^\prime)=-\ri\langle \mathcal{T}_c
h^{}_i(t)h^{\dagger}_i(t^\prime)\rangle$, where $\mathcal{T}_c$ denotes the time-ordering along the contour.  Within DMFT, a local Green's function on the lattice can also be calculated as an impurity Green's function that evolves within the dynamical mean field. Hence, we can also write the core-hole propagator as
\begin{equation}
G_h^c(t,t^\prime)=-\frac{\ri}{\mathcal{Z}_{\textrm{imp}}} \Tr \left \{ \re^{-\beta\mathcal{H}^{\textrm{loc}}}\mathcal{T}_c
\exp \left [-\ri\int_c \rd\bar t\int_c \rd\bar t^\prime d^\dagger(\bar t)\lambda_c(\bar t,\bar t^\prime)d(\bar t^\prime)\right ]
h(t)h^\dagger(t^\prime)\right \}.
\label{eq: g_h}
\end{equation}
Here, $\mathcal{H}^{\textrm{loc}}$ is the local Hamiltonian evaluated at one lattice site and $\mathcal{Z}_\textrm{imp}$ is the impurity partition function calculated for an impurity evolving with respect to the local Hamiltonian and the dynamical mean field evolution operator, given by the exponential function above.  In the next section, we briefly discuss the strategies that can be used to calculate the core-hole propagator.

The XANES signal arises when the X-ray is tuned just slightly above the absorption edge. So, a core hole is created and the photoelectron goes to an unoccupied state in the conduction band.  When this occurs, a photon is taken out of the X-ray beam, and we measure this absorption by counting the number of photons that pass through the sample and comparing this with the number that pass when the sample is not there. Viewed in this way, the XANES experiment is similar to an optical conductivity experiment, except that in the latter, it is an optical photon that causes an excitation of an electron from an occupied to an unoccupied state~\cite{mahan}. Nevertheless, the similarity between these two processes tells us that we can find the XANES signal by evaluating the imaginary part of the Fourier transform of the appropriate current-current correlation function. Here is how we find the corresponding current operator for a XANES experiment. First, we assume that the core-hole and the conduction electron are created at the same site, and for the moment, we assume the core hole will only create itinerant electrons, not localized electrons. Hence, the transition is to one band only.  Second, the photons have momentum $\mathbf{q}$, frequency $\omega_q=cq$, and a spatial profile for the vector potential that describes the light field given by an amplitude $A$ multiplied by a shape function $f_{\mathbf{q}}(\mathbf{r})$.  We define the dipole matrix element at the lattice site $i$ via
\begin{equation}
M_{\mathbf{q}}(i)=\langle 0|d_i^\dagger \frac{\ri e\hbar}{m_ec} f_{\mathbf{q}}(\mathbf{r}) \re^{\ri\mathbf{q}\cdot\mathbf{r}}\mathbf{e}_{\mathbf{q}}\cdot\mathbf{\nabla} h_i^\dagger|0\rangle,
\label{eq: matrix_element}
\end{equation}
which is the matrix element between a core-hole state $h^\dagger_i|0\rangle$ and an electronic state in the conduction band $d^{}_i|0\rangle$, both of which need to be represented by their corresponding wavefunctions in coordinate space. Here, $\mathbf{e}_\mathbf{q}$ is the polarization of the light field. In a translationally invariant system, the matrix element is independent of the lattice site $i$, so the current operator becomes
\begin{equation}
j_{\mathbf{q}}=\re^{-\ri\omega_{\mathbf{q}}t}M_{\mathbf{q}}d^{}_i h^\dagger_i+\re^{\ri\omega_{\mathbf{q}}t} M^*_{\mathbf{q}}h^{}_i d^\dagger_i\, .
\label{eq: current}
\end{equation}
The response of the current is given by the current-current correlation function.  Noting that for XANES we have no core hole in the initial state, we find that the XANES spectra become
\begin{equation}
\chi^{\textrm{XANES}}(\omega_{\mathbf{q}})=-\frac{2|M_{\mathbf{q}}|^2}{\hbar}\Im\int_0^\infty \rd t \re^{\ri\omega_{\mathbf{q}} t} \left\langle d^\dagger(t)h(t)d(0)h^\dagger(0)\right\rangle,
\label{eq: xanes}
\end{equation}
where the local thermodynamic expectation value can be evaluated for the impurity problem, just like the core-hole propagator (implying we need to introduce the dynamical mean-field evolution operator, etc.).

The XES spectra are evaluated in a similar way, but this time there is a core hole present in the initial state, so different terms from the current-current correlation function contribute and we find
\begin{equation}
\chi^{\textrm{XES}}(\omega_{\mathbf{q}})=-\frac{2|M_{\mathbf{q}}|^2}{\hbar} \Im \int_0^\infty \rd t \re^{-\ri\omega_{\mathbf{q}} t} \left\langle d(t)h^\dagger(t)d^\dagger(0)h(0)\right\rangle.
\label{eq: xes}
\end{equation}

In the remainder of this work, we will assume that the matrix element $M_{\mathbf{q}}$ is independent of momentum (and hence independent of the frequency $\omega_{\mathbf{q}}$).

\section{Formalism to evaluate the XANES correlation function}

We will focus primarily on showing the details of an exact evaluation of the XANES correlation function. The XPS function can be evaluated using the techniques similar to those developed earlier for the $f$-electron
spectral function, which employ either a matrix-based formalism or a Wiener-Hopf sum-equation approach~\cite{brandt_urbanek,freericks_turkowski_zlatic,shvaika_wh}. We are preparing a separate publication on this problem~\cite{nandan} which also examines
the numerical renormalization group approach~\cite{anders} for the core-hole propagator.

To calculate the XANES signal, we need to evaluate the four-point function that is defined by the functional derivative
of the core-hole Green's function with respect to the dynamical mean field
\begin{equation}
\frac{\delta G_h^c(t,t^\prime)}{\delta \lambda_c(t_2,t_1)}=
\left\langle \mathcal{T}_ch(t)h^\dagger(t^\prime)d(t_1)d^\dagger(t_2)\right\rangle+G_h^c(t,t^\prime)G^c_d(t_1,t_2),
\label{eq: functional_deriv}
\end{equation}
where the symbol $c$ denotes the contour-ordered quantities and the itinerant hole Green's function $G_d^c$ is defined
analogous to equation~(\ref{eq: g_h}) but with $h(t)h^\dagger(t^\prime)$ replaced by $d(t)d^\dagger(t^\prime)$; this Green's function can also be easily constructed from the retarded Green's function and the Fermi-Dirac distribution as a function of frequency because it is in equilibrium, but we do not give the details here. We need to evaluate the four-point function for $t^\prime=0$, $t_1=0$ and $t_2=t$ before we take the Fourier transform.

Calculation of this four-point function is straightforward, but tedious.  To do so, requires us to define a fair amount of symbols before we can get to the final formula.  In doing so, we  closely follow the work in reference~\cite{shvaika_wh}, which shows how to represent these objects in terms of determinants of Toeplitz matrices, which can either be evaluated directly numerically, or can be solved in the large time limit asymptotically using Szeg\"o's theorem~\cite{szego,szego1,szego2,szego3} within the Wiener-Hopf approach~\cite{wiener_hopf,krein} as developed by McCoy and Wu~\cite{mccoy_wu}. We do not provide details of those approaches here.

We start by defining the fermionic Matsubara frequencies via $\ri\omega_n=\ri\pi T(2n+1)$.  Then, the impurity partition function can be represented as $\mathcal{Z}_{\textrm{imp}}=\mathcal{Z}_{00}+\mathcal{Z}_{01}+\mathcal{Z}_{10}+\mathcal{Z}_{11}$, with
\begin{equation}
\mathcal{Z}_{n_hn_f}=\re^{-\beta (E_fn_f+E_hn_h+Q_fn_fn_h)}\left [ 1+\re^{\beta(\mu-Un_f-Q_dn_h)}\right ]
\prod_{m=-\infty}^{\infty}\left [1-\frac{\lambda(\ri\omega_m)}{\ri\omega_m+\mu-Un_f-Q_dn_h}\right ]
\label{eq: z_def}
\end{equation}
with $n_f=0,1$ and $n_h=0,1$ the two possible fillings for each localized particle. Here, $\lambda(\ri\omega_m)$ is the dynamical mean field evaluated at the Matsubara frequencies [which can be found by using the spectral formula given $\lambda(\omega)$]. Alternatively, these different partition functions can be defined over the contour via
\begin{equation}
\mathcal{Z}_{n_hn_f}=\Tr\left \{\re^{-\beta H^{\textrm{loc}}(n_h,n_f)}\mathcal{T}_c
\exp \left [-\ri\int_c \rd\bar t\int_c \rd\bar t^\prime d^\dagger(\bar t)\lambda_c(\bar t,\bar t^\prime)d(\bar t^\prime)\right ]
\right \},
\label{eq: z_def2}
\end{equation}
so that the functional derivative satisfies
\begin{align}
\label{eq: z_funct_deriv}
\frac{\delta \mathcal{Z}_{n_hn_f}}{\delta\lambda_c(t_2,t_1)}&=\ri\Tr\left \{\re^{-\beta H^{\textrm{loc}}(n_h,n_f)}\mathcal{T}_c
\exp \left [-\ri\int_cd\bar t\int_cd\bar t^\prime d^\dagger(\bar t)\lambda_c(\bar t,\bar t^\prime)d(\bar t^\prime)\right ]
d(t_1)d^\dagger(t_2)\right \}\\
&=-\mathcal{Z}_{n_hn_f}G_{0d,n_hn_f}(t_1,t_2),\nonumber
\end{align}
where $G_{0d,n_hn_f}$ is the effective medium projected onto the sectors of fixed $n_h$ and $n_f$, and is defined in the equation above.

Using these definitions, one has, for example,  $\langle f^\dagger f\rangle=(\mathcal{Z}_{01}+\mathcal{Z}_{11})/\mathcal{Z}_{\textrm{imp}}$. Since $\beta E_h\rightarrow\infty$ for any reasonable temperature, we usually have $\mathcal{Z}_{1n_f}\rightarrow 0$ ($T$ is typically less than 0.1~eV, while $E_h$ is typically larger than 100~eV).

Now we are ready to start the evaluation of the functional derivative.  Note that we always have $t\geqslant t^\prime=0$, so we are evaluating the derivative of the greater core-hole Green's function, which has $n_h=0$ in the initial state and the greater core-hole Green's function satisfies
\begin{equation}
G^>_h(t,0)=G^>_{h,00}(t,0)+G^>_{h,01}(t,0)
\label{eq: g_greater}
\end{equation}
with
\begin{align}
G^>_{h,00}(t,0)&=-\ri\frac{\mathcal{Z}_{00}}{\mathcal{Z}_{\textrm{imp}}}\re^{-\ri E_ht+C_{00}(t)},\label{eq: gh_00}\\
G^>_{h,01}(t,0)&=-\ri\frac{\mathcal{Z}_{01}}{\mathcal{Z}_{\textrm{imp}}}\re^{-\ri (E_h+Q_f)t+C_{01}(t)},
\label{eq: gh_01}
\end{align}
and
\begin{equation}
C_{n_hn_f}(t)=\ln \det\nolimits_c \left [ \mathbb{I}^c-\mathbb{Q}^c_d(t,0)\mathbb{G}_{0d,n_hn_f}\right ].
\end{equation}
The determinant is taken over the contour of the continuous matrix operator defined inside the brackets.  Here,
$\mathbb{I}^c$ is the identity matrix on the contour, $\mathbb{Q}_d^c(t,0)$ is a diagonal matrix equal to  $Q_d$ on the top branch of the contour only [$\mathbb{Q}_d^c(t,0)=Q_d \mathbb{I}^{[0,t]}$, with $\mathbb{I}^{[0,t]}$ equal to the identity matrix for times on the upper branch, and equal to zero otherwise], and $\mathbb{G}_{0d,n_hn_f}$ is the continuous matrix operator for the
effective medium projected onto the sectors of fixed $n_h$ and $n_f$. As it was already mentioned above, the greater core-hole Green's function (\ref{eq: g_greater}) defines the XPS signal.

The functional derivative of $\mathcal{Z}_{n_hn_f}$ is in equation~(\ref{eq: z_funct_deriv}).  We also need the functional derivative of $C_{n_hn_f}(t)$, which satisfies
\begin{equation}
\frac{\delta C_{n_hn_f}(t)}{\delta\lambda_c(t_2,t_1)}=-\Tr_c\left \{ \left [ \mathbb{I}^c-\mathbb{Q}_d^c(t,0)\mathbb{G}_{0d,n_hn_f}\right ]^{-1}\mathbb{Q}_d^c(t,0)\frac{\delta\mathbb{G}_{0d,n_hn_f}}{\delta\lambda_c(t_2,t_1)}\right \}.
\label{eq: c_funct_deriv}
\end{equation}
The functional derivative of the projected effective medium is straightforward to calculate and it yields
\begin{equation}
\frac{\delta G_{0d,n_hn_f}(\bar t,\bar t^\prime)}{\delta\lambda_c(t_2,t_1)}=G_{0d,n_hn_f}(\bar t,t_2)G_{0d,n_hn_f}(t_1,\bar t^\prime).
\label{eq: g0_funct_deriv}
\end{equation}
Substituting equation~(\ref{eq: g0_funct_deriv}) into equation~(\ref{eq: c_funct_deriv}) yields
\begin{align}
\frac{\delta C_{n_hn_f}(t)}{\delta\lambda_c(t_2,t_1)}&=-\left \{ \mathbb{G}_{0d,n_hn_f}\left [ \mathbb{I}^c-\mathbb{Q}_d^c(t,0)\mathbb{G}_{0d,n_hn_f}\right ]^{-1}\mathbb{Q}_d^c(t,0)\mathbb{G}_{0d,n_hn_f}\right \}_{t_1t_2}\nonumber\\
&=G_{0d,n_hn_f}(t_1,t_2)-G^Q_{0d,n_hn_f}(t_1,t_2),
\label{eq: c_funct_deriv2}
\end{align}
where we have defined
\begin{equation}
G^Q_{0d,n_hn_f}(t_1,t_2)=\left \{ \mathbb{G}_{0d,n_hn_f}\left [ \mathbb{I}^c-\mathbb{Q}_d^c(t,0)\mathbb{G}_{0d,n_hn_f}\right ]^{-1}\right \}_{t_1t_2},
\label{eq: gq_def}
\end{equation}
which satisfies the equation
\begin{equation}
G^Q_{0d,n_hn_f}(t_1,t_2)=G_{0d,n_hn_f}(t_1,t_2)+Q_d\int_0^t \rd\bar t G_{0d,n_hn_f}(t_1,\bar t)G_{0d,n_hn_f}^Q(\bar t,t_2).
\label{eq: gq_eqn}
\end{equation}
Plugging this result into equation~(\ref{eq: functional_deriv}) and taking the limits $t_1=t^\prime=0$ and $t_2=t$ gives our final result
\begin{equation}
\chi^{\textrm{XANES}}(t)=\left\langle\mathcal{T}_c h(t)d^\dagger(t)d(0)h^\dagger(0)\right\rangle=
-\ri G_{h,00}^>(t,0)G^Q_{0d,00}(0,t)-\ri G^>_{h,01}(t,0)G^Q_{0d,01}(0,t).
\label{eq: xanes_final}
\end{equation}
What is remarkable about this formula is that once one has projected onto the different static particle number sectors, the expectation values can be evaluated by factorization similar to Wick's theorem, even though we make no such assumption and the above result is exact. Since the core-hole propagator displays the phenomena of the orthogonality catastrophe, where, in the limit as $T\rightarrow 0$, the spectral function diverges with a power law that depends on the coupling strength~\cite{mahan,nozieres}, we expect the same behavior to occur in the XANES (and XES) signals because they involve matrix determinants that are very similar to the XPS signal and should have the similar asymptotic behavior.

In order to calculate this expression, we solve equation~(\ref{eq: gq_eqn}) via Cramer's rule, and then manipulate the determinants to get a uniform expression.  We start by discretizing the interval $[0,t]$ with $N$ time slices ($t_1=0$ and $t_N=t$), with a step size of $\Delta t=t/(N-1)$. Then, we find that we can approximate the XANES correlation function by taking determinants of discrete $(N-1)\times(N-1)$ matrices (which can be scaled to $\Delta t\rightarrow 0$ by repeating with different $\Delta t$ values). The formula is
\begin{align}
 \chi^{\textrm{XANES}}(t)&=-\frac{\mathcal{Z}_{00}}{\mathcal{Z}_{\textrm{imp}}} \re^{-\ri E_ht}\\
&\times
 \det\left [\begin{array}{cccc}
 G_{0d,00}(t_1-t_N) & -\Delta t Q_d G_{0d,00}(t_1-t_2) & \cdots & -\Delta t Q_d G_{0d,00}(t_1-t_{N-1}) \\
 G_{0d,00}(t_2-t_N) & 1-\Delta t Q_d G_{0d,00}(0) & \cdots & -\Delta t Q_d G_{0d,00}(t_2-t_{N-1}) \\
 \vdots & \vdots & \ddots & \vdots \\
 G_{0d,00}(\Delta t) & -\Delta t Q_d G_{0d,00}(t_{N-1}-t_2) & \cdots & 1-\Delta t Q_d G_{0d,00}(0)
 \end{array} \right ]\nonumber
 \\
      &-\frac{\mathcal{Z}_{01}}{\mathcal{Z}_{\textrm{imp}}} \re^{-\ri(E_h+Q_f)t}\nonumber\\
&\times
\det\left [\begin{array}{cccc}
G_{0d,01}(t_1-t_N) & -\Delta t Q_d G_{0d,01}(t_1-t_2) & \cdots & -\Delta t Q_d G_{0d,01}(t_1-t_{N-1}) \\
G_{0d,01}(t_2-t_N) & 1-\Delta t Q_d G_{0d,01}(0) & \cdots & -\Delta t Q_d G_{0d,01}(t_2-t_{N-1}) \\
\vdots & \vdots & \ddots & \vdots \\
G_{0d01}(\Delta t) & -\Delta t Q_d G_{0d,01}(t_{N-1}-t_2) & \cdots & 1-\Delta t Q_d G_{0d,01}(0)
\end{array} \right ].
    \nonumber
\end{align}
Note that since each time $t$ is independent of each other, this calculation can be easily parallelized; the $t$ dependence in the determinant comes from $\mathbb{Q}_d$ which is nonzero only on the interval $[0,t]$ and determines the size of the matrix.
The calculation for the XES signal is similar, but the details will not be given here.

\section{Numerical results}

For the numerical work, we will examine the Falicov-Kimball model on a hypercubic lattice at half-filling for the conduction holes and for the localized holes.
We evaluate the XPS and XANES spectra for three different cases: $U=0.5$ which is a diffusive metal, $U=1$ which is an anomalous metal that has a dip in the density of states at the chemical potential, and $U=2$ which is in the Mott-insulating phase (due to space constraints we do not include XES results).  We choose the core-hole interactions to satisfy $Q_d=Q_f=2$.  Both the XPS in equation~(\ref{eq: g_greater}) and XANES in equation~(\ref{eq: xanes_final}) responses contain two contributions (groups of peaks), one from the sites with empty $f$-states and another from the occupied ones, separated by the interaction energy $Q_f$ and are represented by the blue dashed and red dot-dashed lines, respectively, in the figures below.

\begin{figure}[htbp]
\centering
\includegraphics[width=0.45\columnwidth,clip]{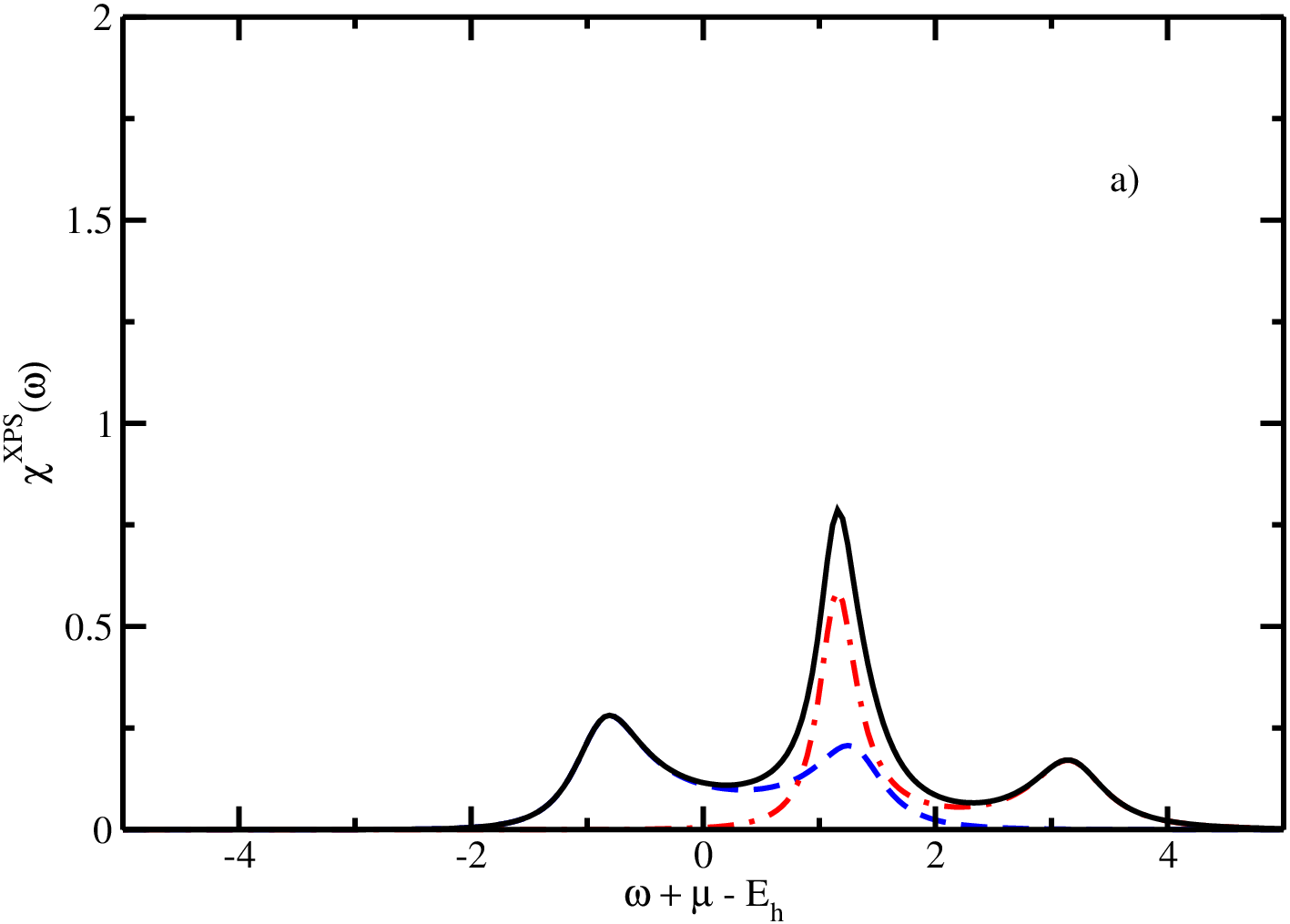}\qquad%
\includegraphics[width=0.45\columnwidth,clip]{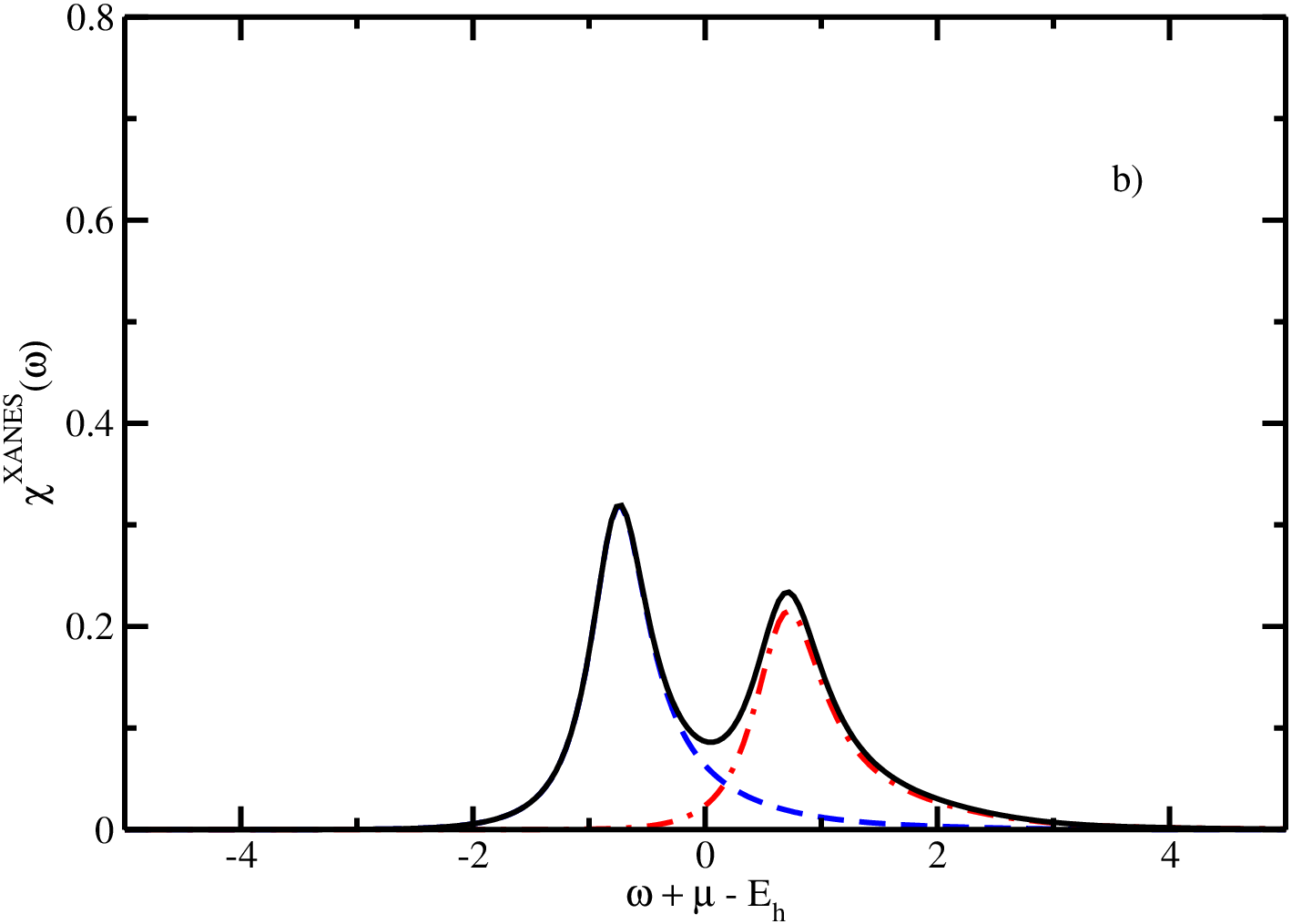}\\ [1ex]
\centering
\includegraphics[width=0.45\columnwidth,clip]{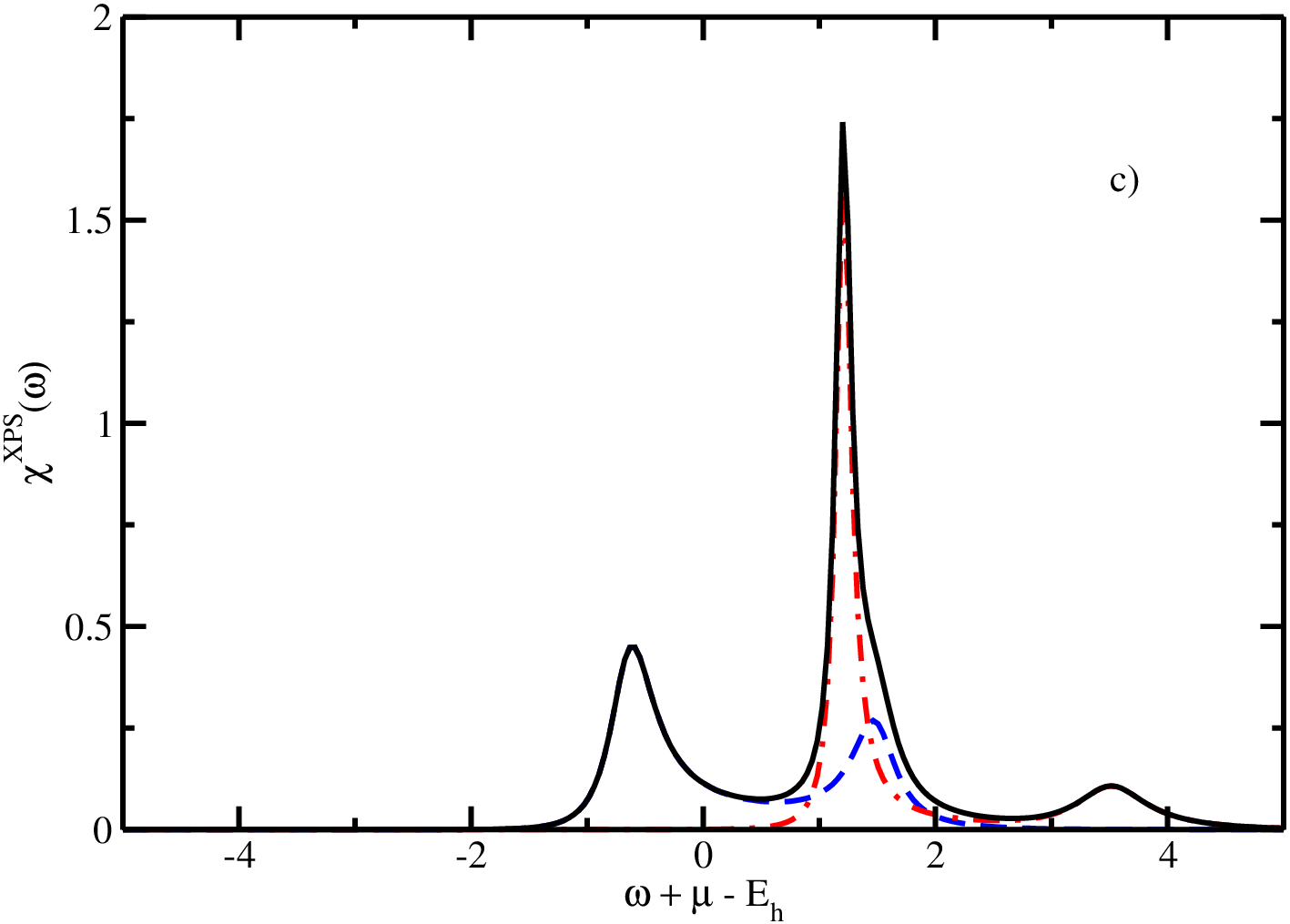}\qquad%
\includegraphics[width=0.45\columnwidth,clip]{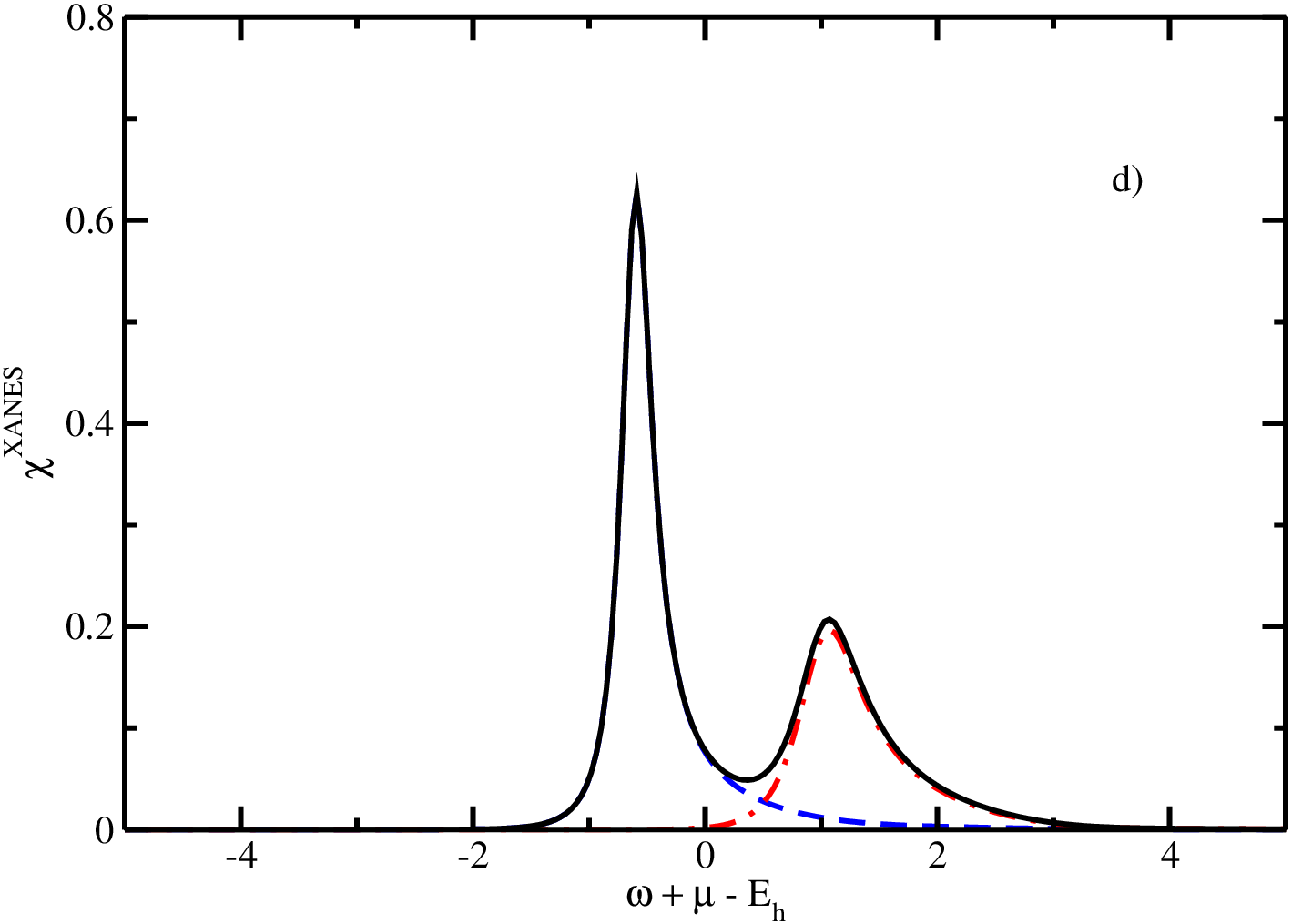}
\caption{(Color online) XPS (a, c) and XANES (b, d) spectra for $U=0.5$, $Q_d=Q_f=2$, and $T=1$ (a, b) and $T=0.3$ (c, d). The black solid line gives the total spectra and the blue dashed and red dot-dashed lines correspond to the first and second contributions in the equations~(\protect\ref{eq: g_greater}) and (\protect\ref{eq: xanes_final}), respectively.}
\label{fig: dif_met}
\bigskip
\centering
\includegraphics[width=0.45\columnwidth,clip]{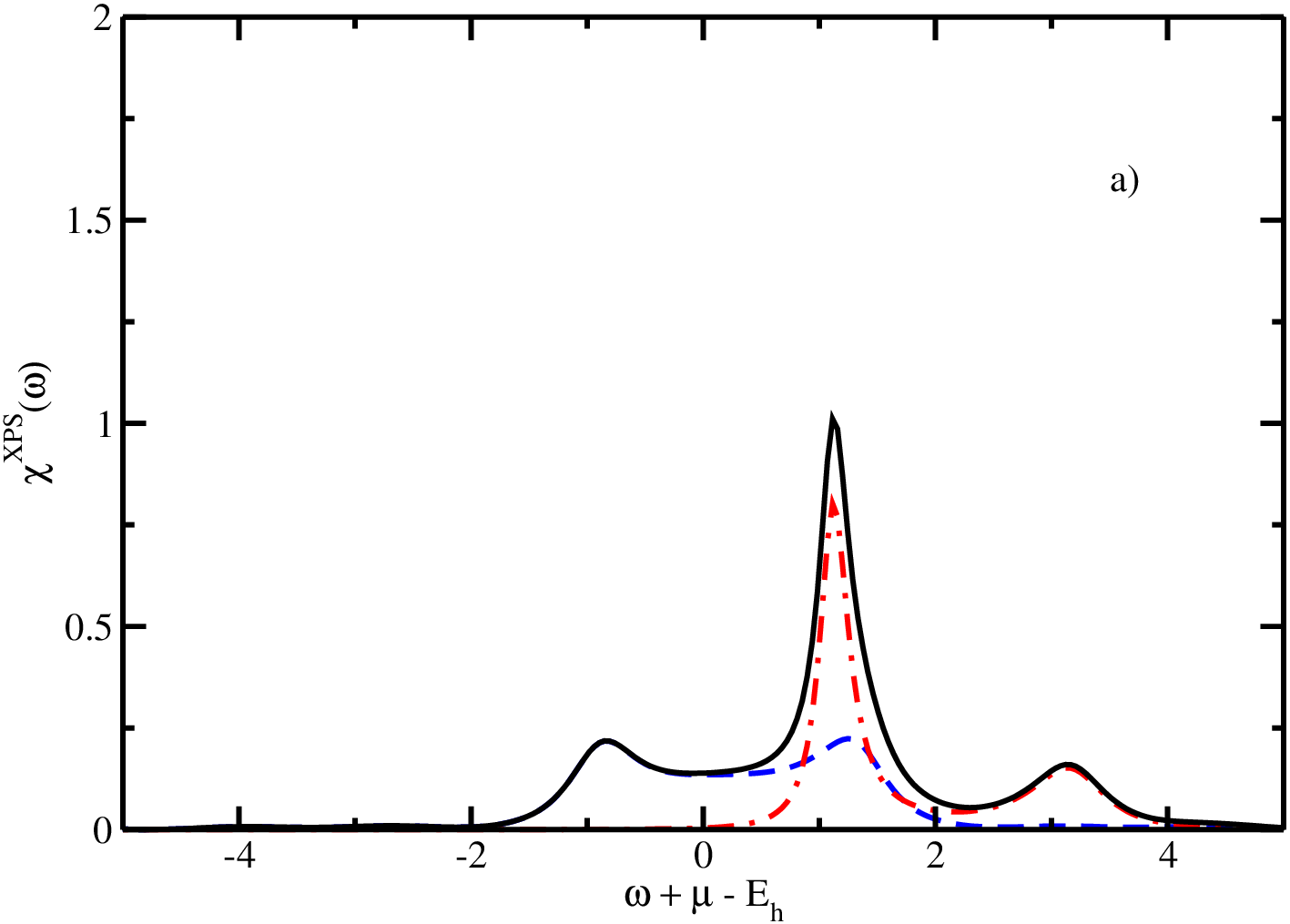}\qquad%
\includegraphics[width=0.45\columnwidth,clip]{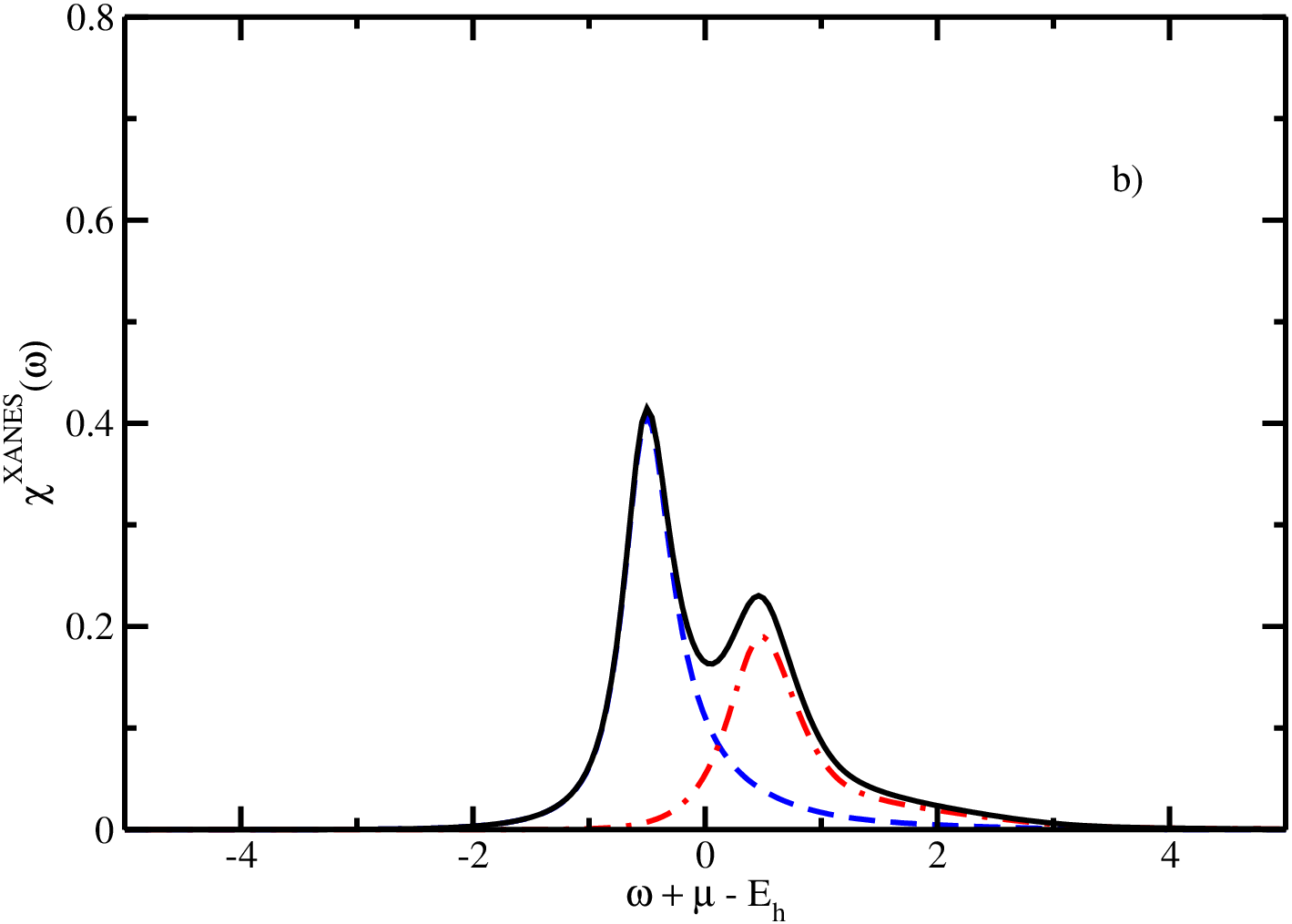} \\ [1ex]
\includegraphics[width=0.45\columnwidth,clip]{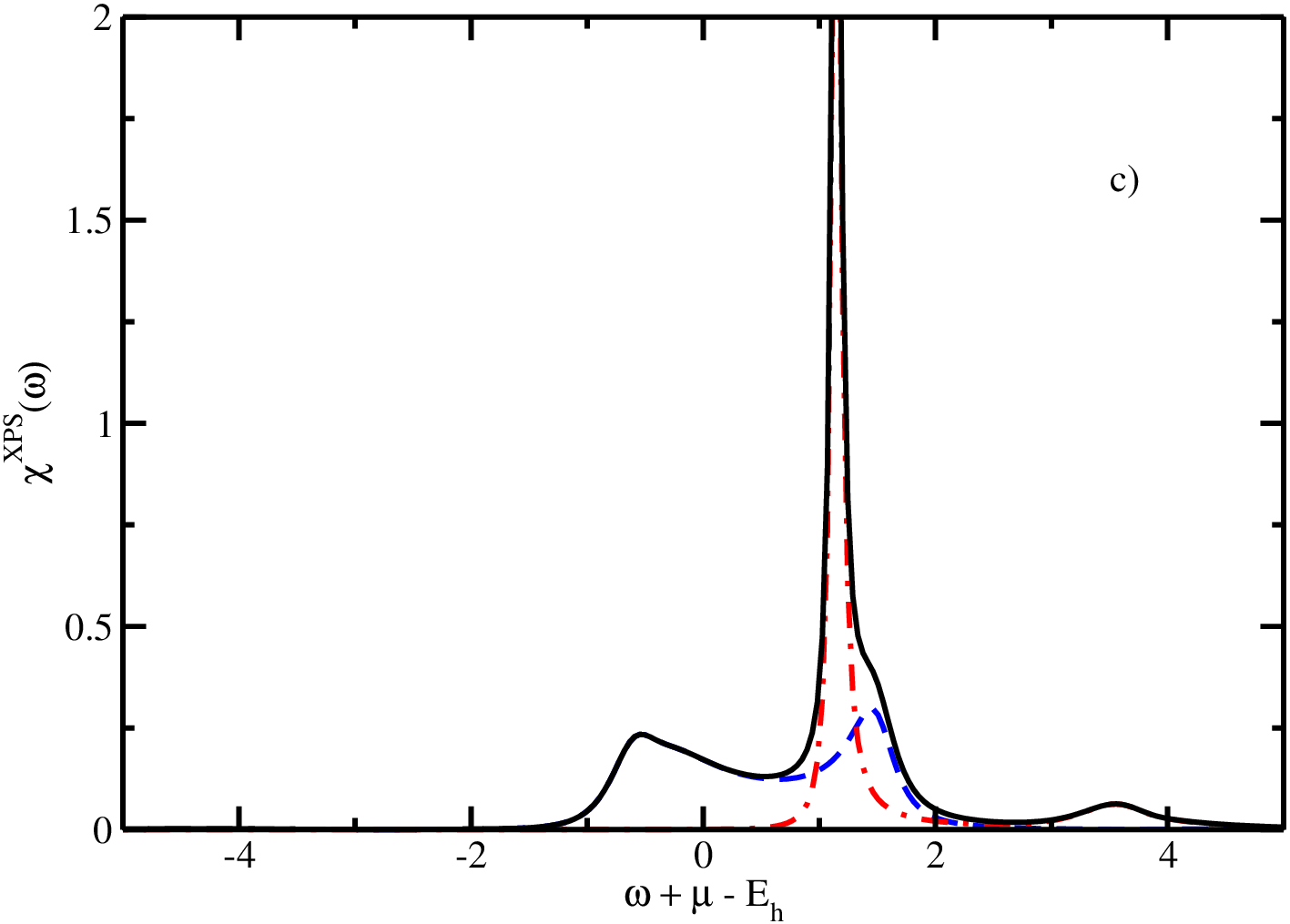}\qquad%
\includegraphics[width=0.45\columnwidth,clip]{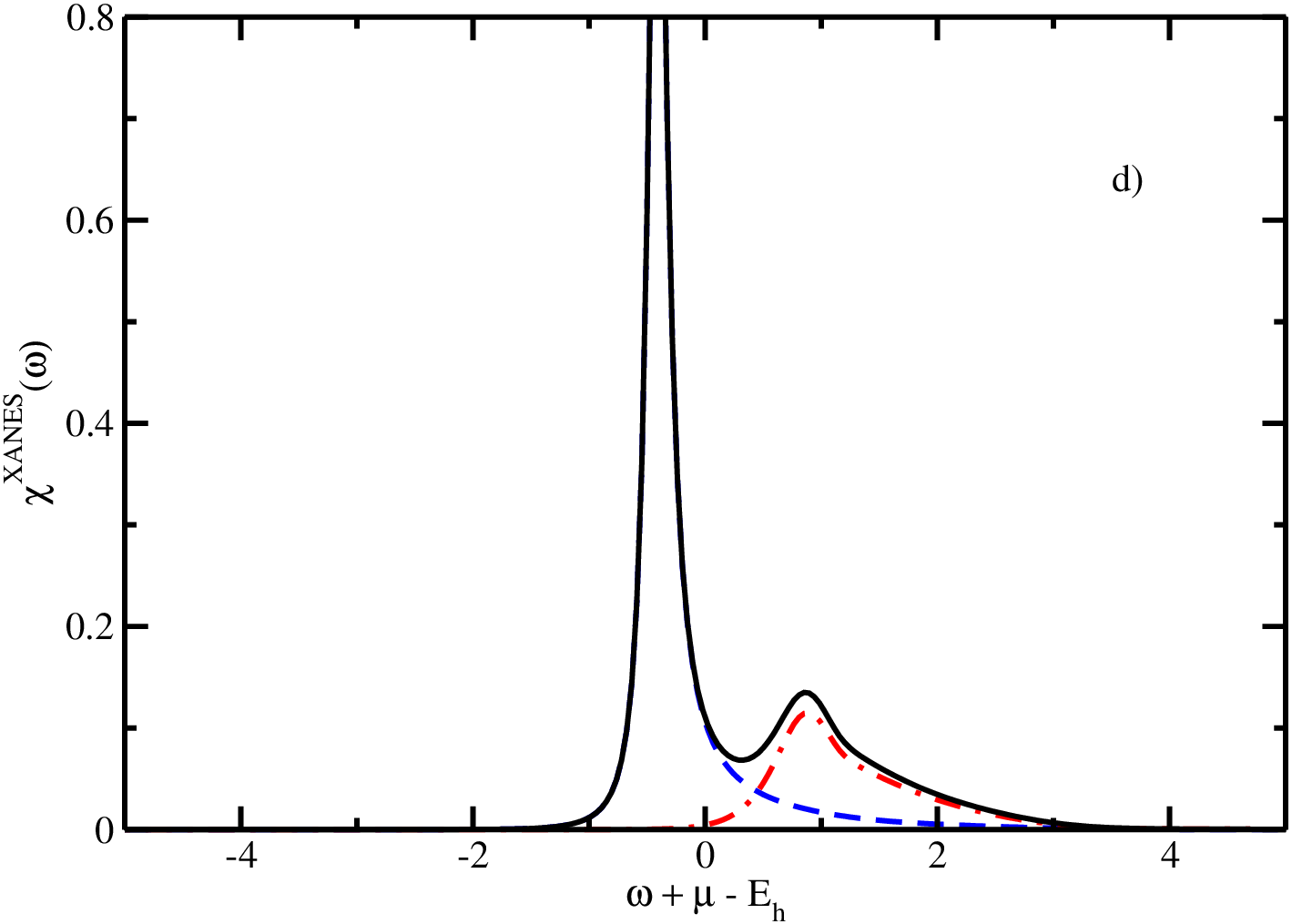}
\caption{(Color online) The same as in figure~\protect\ref{fig: dif_met} but for $U=1$.}
\label{fig: anom_met}
\end{figure}

For the cases of the diffusive (figure~\ref{fig: dif_met}) and anomalous (figure~\ref{fig: anom_met}) metals, each of the contributions to the XPS response contain two peaks separated by the interaction energy $Q_d$. At high temperatures for the case of the diffusive metal (figure~\ref{fig: dif_met}a), we have in total four peaks (with two peaks merged due to the parameter selection $Q_d=Q_f$). When the temperature becomes lower (figure~\ref{fig: dif_met}c), the peaks become more pronounced with an almost Lorentzian profile and one of them becomes very sharp and at zero temperature transforms into the power law singularity of the Anderson orthogonality catastrophe~\cite{nozieres}. For the case of the anomalous metal (figure~\ref{fig: anom_met}), the XPS response is almost the same with the appearance of additional shoulders at the peaks and now the edge singularity is much narrower and sharper due to the decrease of the density of states at the chemical potential level, but it is still a power law divergence at zero temperature.
The XANES response for these cases of the diffusive and anomalous metal contains two peaks, one of which transforms into a power law singularity at zero temperature. In other aspects, the behavior is similar to the XPS response.

\begin{figure}[!t]
\centering
\includegraphics[width=0.45\columnwidth,clip]{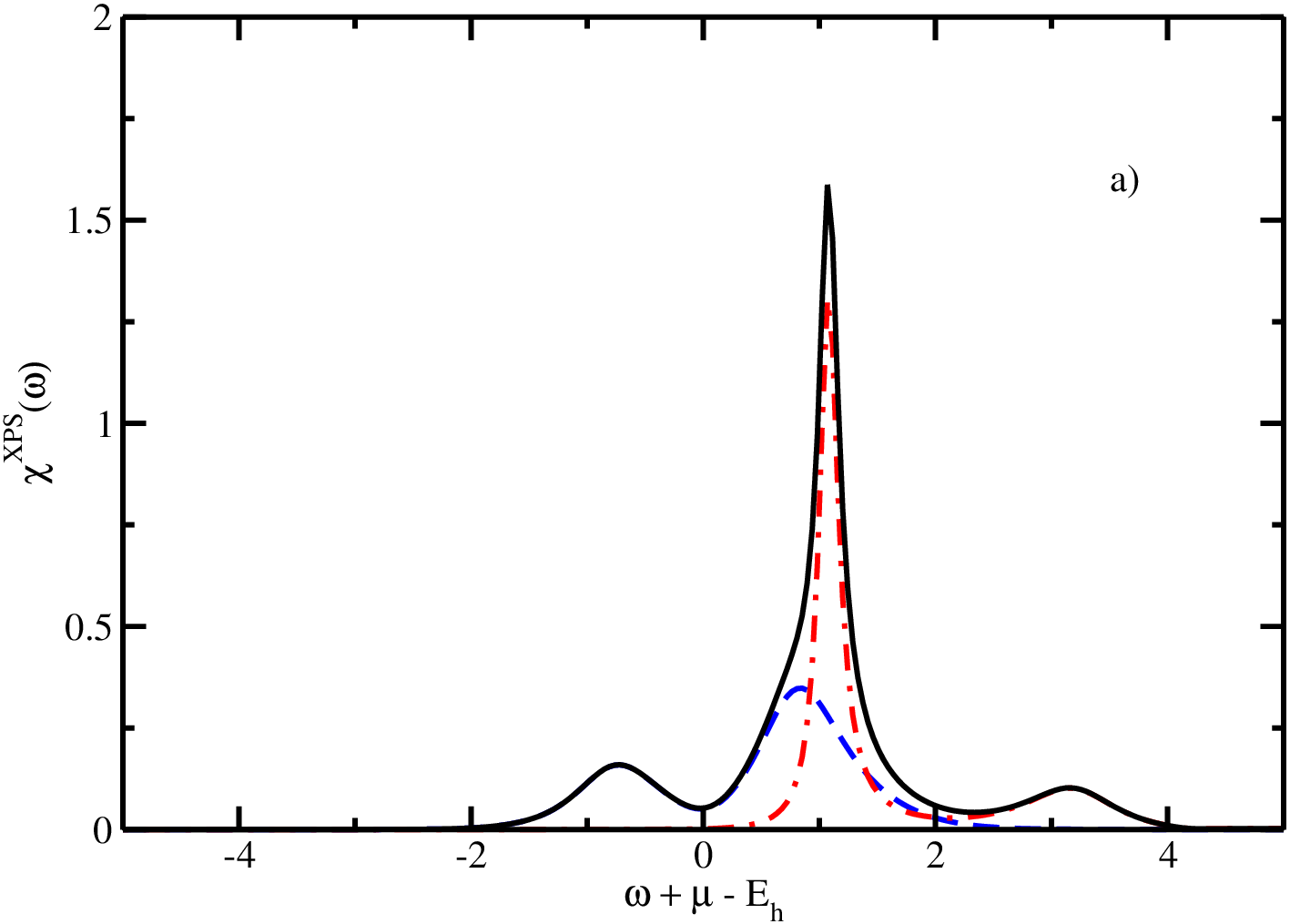}\qquad%
\includegraphics[width=0.45\columnwidth,clip]{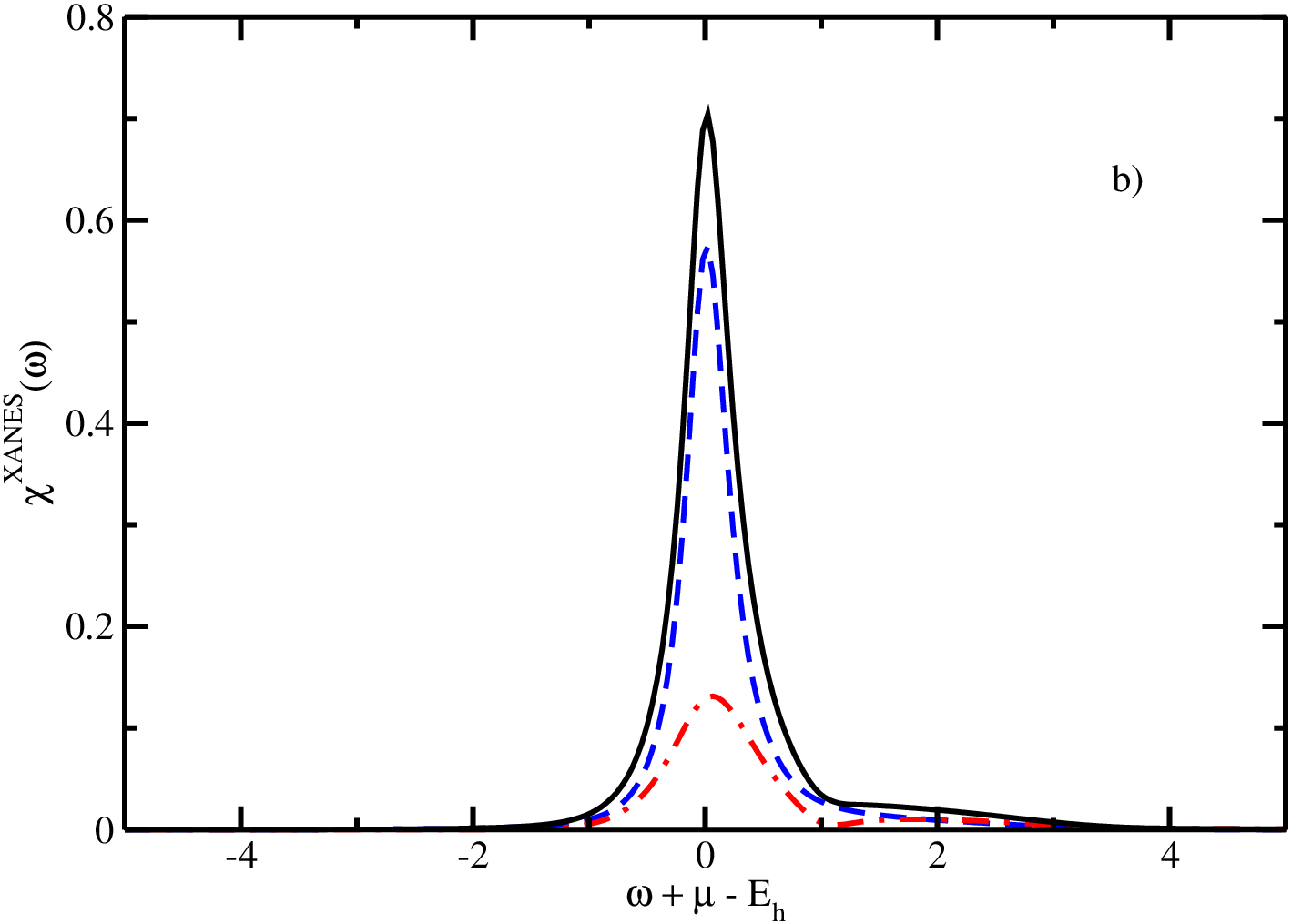}\\ [1ex]
\centering
\includegraphics[width=0.45\columnwidth,clip]{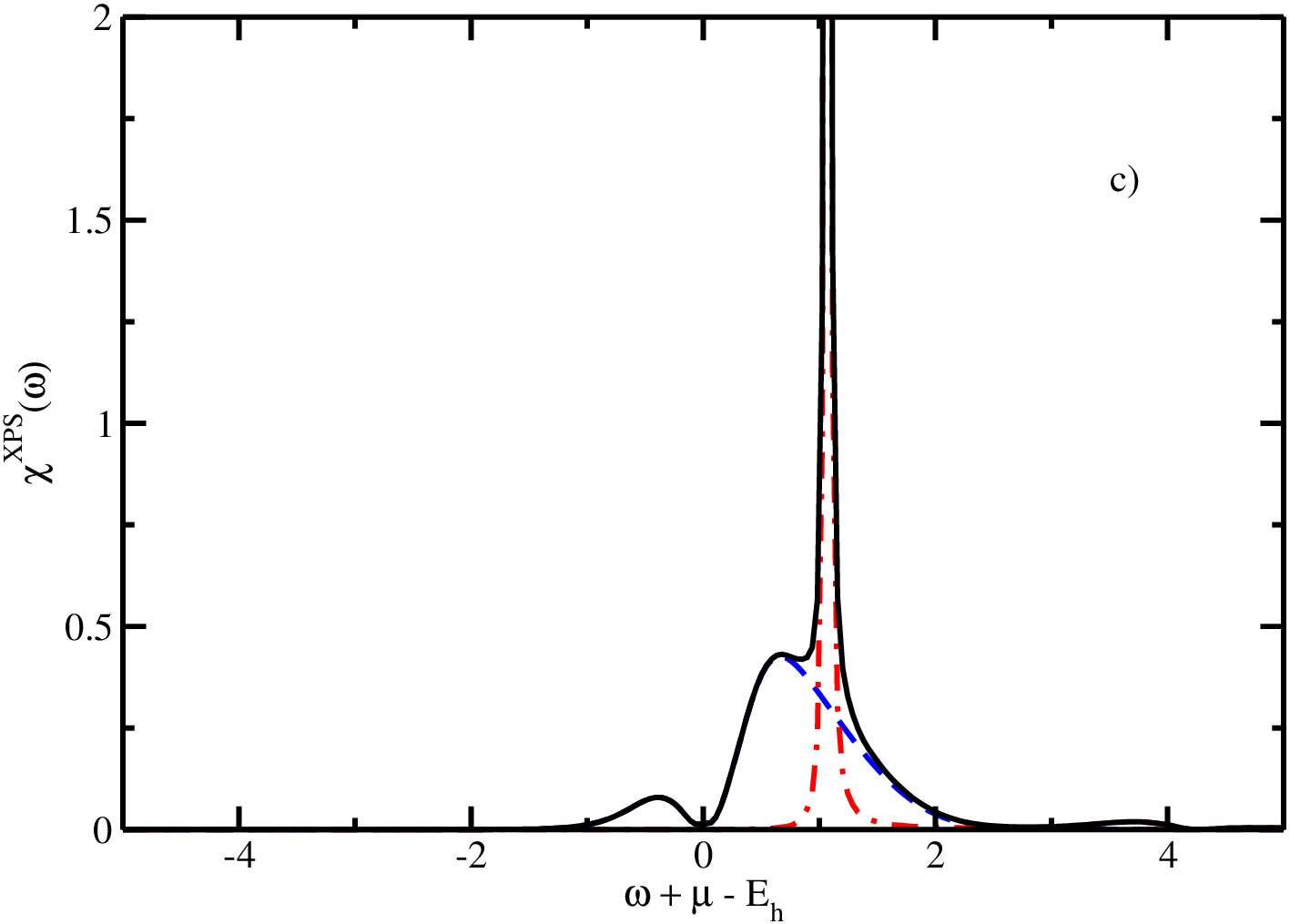}\qquad%
\includegraphics[width=0.45\columnwidth,clip]{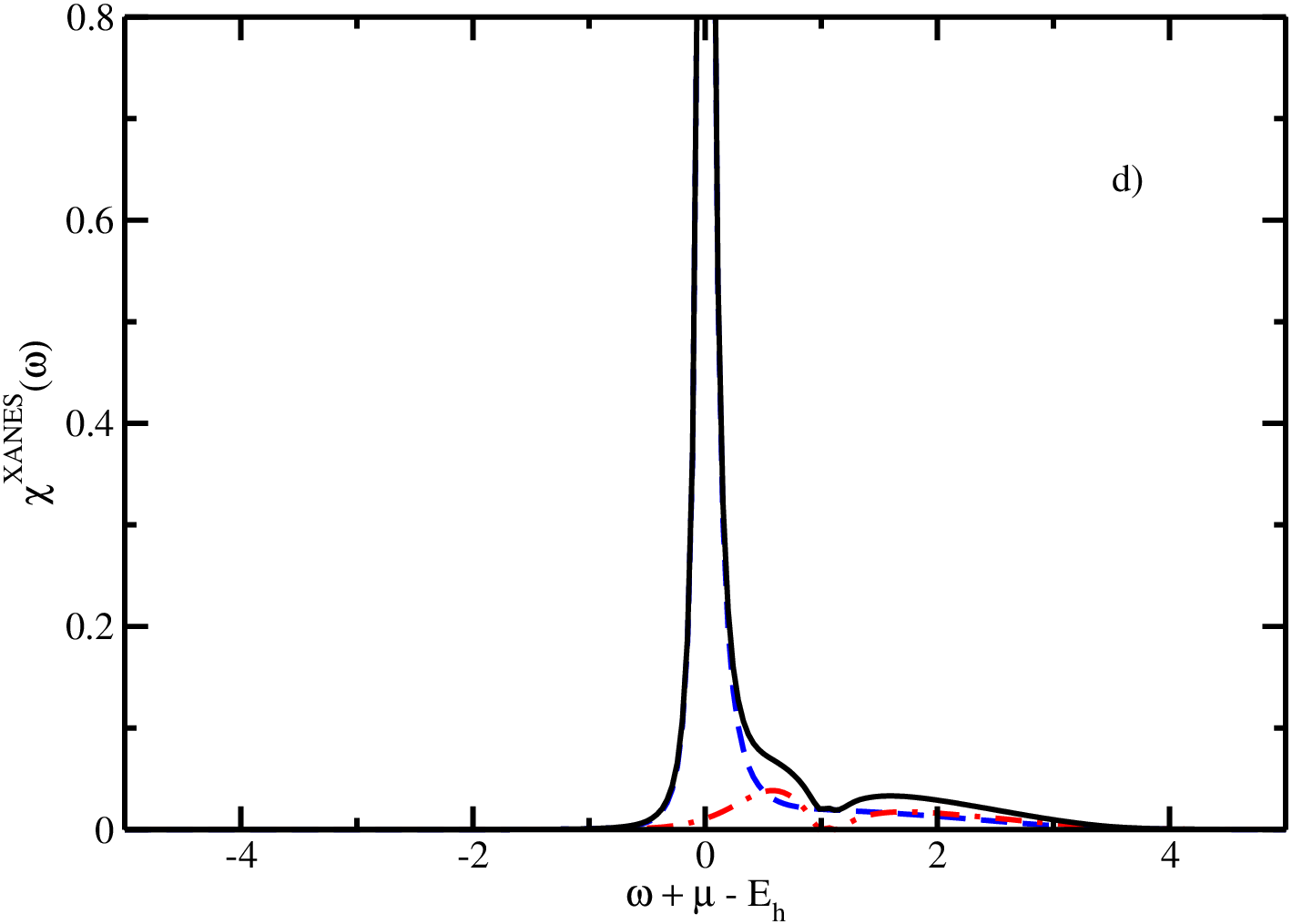}
\caption{(Color online) The same as in figure~\protect\ref{fig: dif_met} but for $U=2$.}
\label{fig: ins}
\end{figure}

The behavior of the XPS and XANES spectra in the Mott-insulating case is completely different (figure~\ref{fig: ins}). The XPS response at high temperatures is similar to the one in the metallic phase, but when the temperature becomes lower, the first group of the peaks displays a gapped spectra similar to what is seen in the $f$-particle density of states~\cite{shvaika_wh}. The other group of the peaks display a very narrow Lorentzian feature which changes at zero temperature from a power law singularity to a $\delta$-peak. The XANES response also displays a sharp edge singularity, but now it is accompanied by a near edge structure which corresponds to the gapped Mott-insulator density of states.

\section{Conclusions}

\looseness=-1
In this work, we have shown how to calculate three different X-ray spectroscopies in strongly correlated materials: XPS, XANES, and XES within the Falicov-Kimball model in the limit of large spatial dimensions. In all cases, one can find analytic expressions for these spectra in terms of determinants of continuous matrix operators.  These continuous operators can be discretized and evaluated on a computer using standard LAPACK routines. The XPS and XANES spectra have similar behavior, but the peak locations and weights differ, indicating how they are measuring different processes.  Generically, the XANES signal has fewer peaks, but the strong-coupling satellites often appear with a higher weight. While these results do not incorporate complicated multiplet or spin-orbit coupling effects, which  are often seen in real materials, they do include all of the strong-correlation effects in a simple model, allowing for an exact solution.

A similar formalism to what was developed here can be used to solve for time-resolved XPS, XANES and XES experiments, where the system is pumped to a nonequilibrium state via a high-intensity femtosecond pulse, and then the system is probed with lower amplitude X-rays after some time delay with respect to the pump. The main difference in the formalism is that Green's functions lose their time-translational invariance and need to be solved using nonequilibrium DMFT~\cite{freericks_nedmft}. We will present the results for that work elsewhere. With the advent of newer
high brilliance, short pulse width, X-ray sources (like the linac coherent light source, LCLS), we expect there to be more studies of pump/probe phenomena in strongly correlated materials, and it is possible that probes like XPS, XANES, or XES, can shed light on interesting nonequilibrium dynamics. XPS has already been employed~\cite{rossnagel} to measure the properties of the order parameter of charge-density-wave systems on ultrafast time scales, indicating that we are at the frontier of these types of experiments.

\newpage

\section*{Acknowledgements}
J.K.F. was supported by the Department of Energy under grant number DE-FG02-08ER46542.  The US-Ukraine collaboration
was supported by the Department of Energy under grant number DE-SC0007091.
J.K.F. also acknowledges support from the McDevitt bequest at Georgetown.

\vspace{-5mm}

\ukrainianpart

\title{Точний розв'язок теорії динамічного середнього поля \\ для різноманітних дослідів з X-променями \\ в моделі Фалікова-Кімбала}
\author{А.М. Швайка\refaddr{label1}, Дж.К. Фрiрiкс\refaddr{label2}}
\addresses{
\addr{label1} Iнститут фiзики конденсованих систем НАН України, вул. Свєнцiцького 1, 79011 Львiв, Україна
\addr{label2} Фiзичний факультет, Унiверситет Джорджтауну, Вашингтон, округ Колумбiя 20057, США
}
%
%
%

\makeukrtitle

\begin{abstract}
\tolerance=3000%
Досліджуються рентґенівські спектри фотоелектронної емісії (XPS), передкрайового поглинання (XANES) та фотоемісії (XES) для моделі Фалікова-Кімбала, використовуючи точні розв'язки теорії динамічного середнього поля. XPS вимірює пропагатор дірки йонного залишку, XANES~--- поглинання X-променів, коли електрон йонного залишку збуджується в незаповнену електронну зону твердого тіла, і XES~--- спектр випроміненого світла при заповненні електроном наперед створеної X-променями дірки йонного залишку. Всі ці три типи спектрів тісно пов'язані між собою і проявляють особливості типу катастрофи ортогональності при $T=0$. Показано, як ефективно розраховувати такі спектри для скінчених температур, зокрема з детальнішим наголосом на спектри XANES.
\keywords фотоелектронна рентґеноспектроскопія (XPS), рентґеноспектроскопія передкрайового поглинання (XANES), теорія динамічного середнього поля, модель Фалікова-Кімбала

\end{abstract}


\vspace{-1mm}

\begin{thebibliography}{99}

 \bibitem{nandan_rixs} Pakhira~N., Freericks~J.K., Shvaika~A.M., Phys. Rev. B,
 2012, \textbf{86}, 125103; \doi{10.1103/PhysRevB.86.125103}.

\bibitem{falicov_kimball} Falicov~L.M., Kimball~J.C., Phys. Rev. Lett., 1969, \textbf{22}, 997; \bibdoi{10.1103/PhysRevLett.22.997}.

\bibitem{freericks_review} Freericks~J.K., Zlati\'c~V., Rev. Mod. Phys., 2003, \textbf{75}, 1333; \bibdoi{10.1103/RevModPhys.75.1333}.

\bibitem{freericks_sudden} Freericks~J.K., Krishnamurthy~H.R., Ge~Yizhi, Liu~A.Y., Pruschke~Th.,
Phys. Status Solidi B, 2009, \textbf{246}, 948; \bibdoi{10.1002/pssb.200881555}.

\bibitem{kadanoff_baym} Kadanoff~L.P., Baym~G., {Quantum statistical mechanics}, Benjamin, New York, 1962.

\bibitem{keldysh} Keldysh~L.V., Zh. Eksp. Teor. Fiz., 1964, \textbf{47}, 1945 (in Russian) [Sov. Phys. JETP, 1965, \textbf{20}, 1018].

\bibitem{mahan} Mahan~G.D., Phys. Rev., 1967, \textbf{163}, 612; \bibdoi{10.1103/PhysRev.163.612}.

\bibitem{brandt_urbanek} Brandt~U., Urbanek~M.P., Z. Phys. B: Condens. Matter, 1992, \textbf{89}, 297; \bibdoi{10.1007/BF01318160}.

\bibitem{freericks_turkowski_zlatic} Freericks~J.K., Turkowksi~V.M., Zlati\'c~V., Phys. Rev. B, 2005, \textbf{71}, 115111;
    \bibdoi{10.1103/PhysRevB.71.115111}.

\bibitem{shvaika_wh} Shvaika~A.M., Freericks~J.K., Condens. Matter Phys., 2008, \textbf{11}, 425.

\bibitem{nandan} Pakhira~N., Shvaika~A.M., Freericks~J.K. (unpublished).

\bibitem{anders} Anders~F.B., Czycholl~G., Phys. Rev. B, 2005, \textbf{71}, 125101; \bibdoi{10.1103/PhysRevB.71.125101}.

\bibitem{szego} Szeg\"o~G., Math. Z., 1920, \textbf{6}, 167.

\bibitem{szego1} Szeg\"o~G., Math. Z., 1921, \textbf{9}, 167.

\bibitem{szego2} Szeg\"o~G., Commun. Sem. Math. Univ. Lund, suppl. ded. Marcel Reisz, 1952, 228.

\bibitem{szego3} Grenander~U., Szeg\"o~G., {Toeplitz forms and their applications}, University of California Press, Berkeley and Los Angeles, 1958.

\bibitem{wiener_hopf} Wiener~N., Hopf~E., Sitzber. Preuss. Akad. Wiss. Berlin, Kl. Math. Phys. Tech., 1931, 696.

\bibitem{krein} Krein~M.G., Uspekhi Mat. Nauk, 1958, \textbf{13}, No.~5(83), 3 (in Russian) [Am. Math. Soc. Transl., 1962, \textbf{22}, 163].

\bibitem{mccoy_wu} McCoy~B.M.,Wu~T.T., {The two-dimensional Ising model}, Harvard University Press, Cambridge, MA, 1973.

\bibitem{nozieres} Nozi\`eres~P., De Dominicis~C., Phys. Rev., 1969, \textbf{178}, 1097; \bibdoi{10.1103/PhysRev.178.1097}.

\bibitem{freericks_nedmft}  Freericks~J.K., Turkowski~V.M., Zlati\'c~V., Phys. Rev. Lett., 2006, \textbf{97}, 266408; \bibdoi{10.1103/PhysRevLett.97.266408}.

\bibitem{rossnagel}  Hellmann~S., Beye~M.,  Sohrt~C., Rohwer~T.,  Sorgenfrei~F.,  Redlin~H.,  Kall\"ane~M., Marczynski-B\"uhlow~M.,  Hennies~F., Bauer~M.,  F\"ohlisch~A.,  Kipp~L.,  Wurth~W., Rossnagel~K., Phys. Rev. Lett., 2010, \textbf{105}, 187401; \bibdoi{10.1103/PhysRevLett.105.187401}.

\end{thebibliography}
\end{document}